\documentclass[prl,showpacs,twocolumn,superscriptaddress]{revtex4}
\usepackage{amsmath}
\usepackage{graphicx} 
\usepackage{dcolumn}  
\usepackage{bm}       


\begin{document}

\author{S. Manz}
 \affiliation{Institute of Atomic and Subatomic Physics, TU-Wien, A-1020 Vienna, Austria}

\author{R. B\"ucker}
 \affiliation{Institute of Atomic and Subatomic Physics, TU-Wien, A-1020 Vienna, Austria}
 
 \author{T.~Betz}
  \affiliation{Institute of Atomic and Subatomic Physics, TU-Wien, A-1020 Vienna, Austria}
 
  \author{Ch.~Koller}
  \affiliation{Institute of Atomic and Subatomic Physics, TU-Wien, A-1020 Vienna, Austria}
 
 \author{S.~Hofferberth}
 \affiliation{Department of Physics, Harvard University, Cambridge, MA 02138, USA}

 \author{I.~E. Mazets}
 \affiliation{Institute of Atomic and Subatomic Physics, TU-Wien, A-1020 Vienna, Austria}
 \affiliation{Ioffe Physico-Technical Institute, 194021 St.Petersburg, Russia}
  \affiliation{Wolfgang Pauli Institute, University of Vienna, 1090 Vienna, Austria}
 
  \author{A.~Imambekov}
  \affiliation{Department of Physics and Astronomy, Rice University, Houston, Texas 77005, USA}
    
 \author{E. Demler}
 \affiliation{Department of Physics, Harvard University, Cambridge, MA 02138, USA}
 
 \author{A. Perrin}
  \affiliation{Institute of Atomic and Subatomic Physics, TU-Wien, A-1020 Vienna, Austria}
     \affiliation{Wolfgang Pauli Institute, University of Vienna, 1090 Vienna, Austria}
   
 \author{J. Schmiedmayer}
  \affiliation{Institute of Atomic and Subatomic Physics, TU-Wien, A-1020 Vienna, Austria}
  
 \author{T. Schumm}
  \affiliation{Institute of Atomic and Subatomic Physics, TU-Wien, A-1020 Vienna, Austria}
 \affiliation{Wolfgang Pauli Institute, University of Vienna, 1090 Vienna, Austria} 
 
\title{Two-point density correlations of quasicondensates in free expansion}

\begin{abstract}
We measure the two-point density correlation function of freely expanding quasicondensates in the weakly interacting quasi-one-dimensional (1D) regime. While initially suppressed in the trap, density fluctuations emerge gradually during expansion as a result of initial phase fluctuations present in the trapped quasicondensate. Asymptotically, they are governed by the thermal coherence length of the system. Our measurements take place in an intermediate regime where density correlations are related to near-field diffraction effects and anomalous correlations play an important role. Comparison with a recent theoretical approach described by Imambekov et al. yields good agreement with our experimental results and shows that density correlations can be used for thermometry of quasicondensates. 
\end{abstract}
\date{\today}
\maketitle


Interacting Bose gases confined in strongly elongated traps exhibit unique properties and quantum phases related to the one-dimensional (1D) character of the underlying physics~\cite{Pet_00,Pet_01,Khe_03}. They also represent one of the few complex many-body systems which allow a direct comparison with exact and often analytical theoretical description~\cite{Ame_08}. For such gases, the smooth transition from a fully decoherent system to a finite-size Bose-Einstein condensate is characterized by an intermediate \emph{quasicondensate} regime where density fluctuations are suppressed~\cite{Est_06} whereas axial phonon-like excitations are thermally populated. As a result the gas displays 1D phase fluctuations along its axial direction, strongly affecting its coherence properties. 

Early experimental evidences for this regime have been obtained in expansion experiments where phase fluctuations of the trapped quasicondensate turn into density fluctuations~\cite{Det_01}. Momentum Bragg spectroscopy~\cite{Ric_03} and more recently matter wave interferometry~\cite{Jo_07,Hof_07,Hof_08} have allowed one to investigate in depth the first-order coherence properties of such systems. More involved interferometry schemes have been used to measure the second order correlation function of trapped quasicondensates~\cite{Hel_03}.

\begin{figure}[b]
\begin{center}
\includegraphics{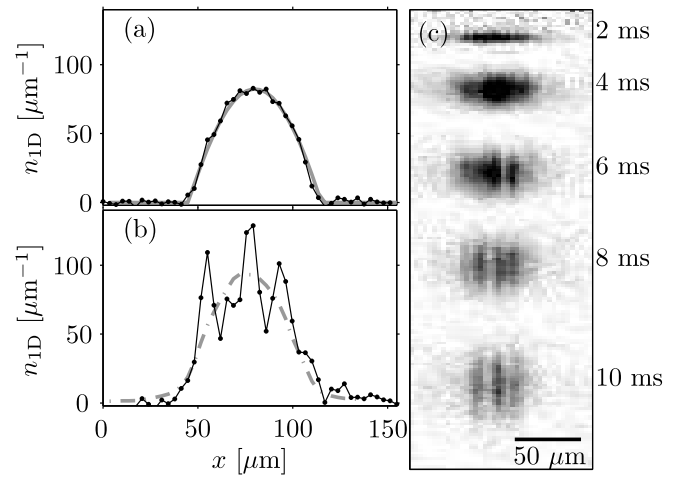}
\caption{(a) Linear density profile for a short expansion time (2\,ms), where density fluctuations have not yet developed, and fitted density profile~\cite{Ger_04}. (b) Linear density profile after 10 ms of free expansion, showing high contrast density fluctuations, where the averaged density profile (gray dashed line) is smooth.
(c) Example images of density fluctuations emerging during free expansion.}
\label{fig:figure1}
\end{center}
\end{figure}

In this letter we report on the experimental measurement of the two-point correlation function of quasicondensates released from a tight atom-chip wave guide. The theoretical description of quasicondensate expansion is motivated by the possibility of a unique mapping of the emerging density correlations back to the trapped system~\cite{Ima_09}. In previous experiments, this was complicated by interaction-induced hydrodynamic effects during expansion~\cite{Det_01,Hel_01}. A complete description of density correlations then requires involved numerical calculations. Our samples are sufficiently deep in the 1D regime so that the radial dynamics can be neglected and the expansion can be considered as collisionless. This allows us to apply the theory of Imambekov et al.~\cite{Ima_09} and directly relate the density correlations in expansion to the correlation length of phase fluctuations present in the trapped system. Our measurements take place in a near-field regime where the shape of the two-point correlation function reflects the transition from the trapped system (suppressed density fluctuations) to the asymptotic case (exponential decay on a length scale given by the in-trap coherence length). Extending on the work with elongated three-dimensional (3D) Bose-Einstein condensates~\cite{Hel_01} we use the amplitude of density fluctuations of quasicondensates for thermometry in a temperature regime where conventional methods based on the visibility of thermal atoms fail.

The starting point for our experiments is a thermal cloud of $^{87}$Rb atoms in the $\left|F=1,m_F=-1\right\rangle$ 
state prepared in a strongly anisotropic Ioffe-Pritchard-type wire trap on a multilayer atom-chip~\cite{Tri_08}. The measured trap frequencies are $\omega_r = 2 \pi \times 2.3$\,kHz in the radial and $\omega_a =2 \pi \times 19$\,Hz in the axial direction. 
We use forced RF evaporation to cool the gas into the quantum degenerate regime, the final temperature is adjusted by limiting the effective trap depth to $\left(1-10\right)\,\hbar \omega_r$. To ensure thermal equilibrium, this situation is maintained for several hundreds of milliseconds. We adjust an additional hold-time to keep an atom number of $\approx$~3500 in all presented experiments, making use of technical losses. This corresponds to a peak atomic density of $n_{\text{1D}}(0)=90\,\mu$m$^{-1}$ and a chemical potential of $\mu/\hbar=2\pi\times1.6$\,kHz$=0.7\,\omega_r$~\cite{Ger_04}. For temperatures $T<\hbar\omega_r/k_B=110\,$nK we realize the regime of weakly interacting quasi-1D Bose gases ($\mu, k_B T \lesssim \hbar\omega_r$) where the radial degrees of motion have negligible influence on the system~\footnote{For $T<20$~nK we would expect a smooth crossover to a finite-size condensate; this is however not accessible in our setup}. In this regime the quasicondensate is characterized by thermal phase fluctuations with a correlation length $\lambda_T = 2 \hbar^2 n_{\rm 1D}/m k_B T$, $m$ being the mass of the atoms. For our experiment $\lambda_T$ ranges from $0.25 R_a$ to $0.5 R_a$ where $2 R_a=64\,\mu$m is the axial size of the gas.

After preparation of the quasicondensate, the confining atom-chip potential is suddenly switched off (within less than $10 \mu s$) and the atomic density is imaged after an expansion time $t_{\text{\rm exp}}$ using standard absorption imaging~\footnote{We find it interesting to note, that (in the quasi-1D regime) important information on the two-point correlation function of quasicondensates can be obtained from standard absorption imaging without the need for single-atom sensitivity or atom shot-noise-limited detection.}. During the switchoff, the expansion changes from hydrodynamic to collisionless on a time scale on the order of $\omega_r^{-1}$. During this time, short-range fluctuations up
to the size of $\delta x = \xi_h \mu/\hbar\omega_r$, $\xi_h = \hbar / \sqrt{m \mu}$ being the healing length, will be smeared, and are not visible in the correlation function~\cite{Ima_09}. Since this length scale is on the order of 100\,nm it cannot be resolved with our imaging system. For the visible density fluctuations we can therefore assume ballistic expansion.

Typical experimental images are depicted in Fig.~\ref{fig:figure1}. Starting from a smooth \emph{in situ} density profile, the atomic density develops density fluctuations in the course of the expansion, reflecting the thermal phase fluctuations initially present in the quasicondensate. In contrast to other work~\cite{Cle_08,Che_08}, density fluctuations in the initial sample (e.g., due to corrugation of the trapping potential) can be neglected here. In expansion, each individual realization exhibits different density fluctuations. The mean image, however, shows a smooth profile.

\begin{figure}[tb]
\begin{center}
\includegraphics{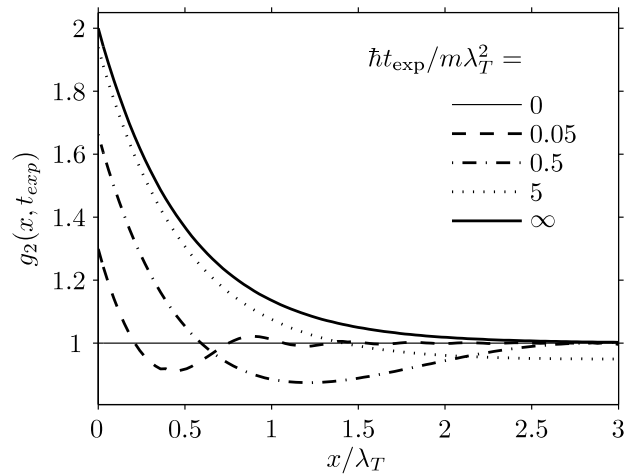}
\caption{Two-point correlation function of an ultracold 1D Bose gas for different times $t_{\text{\rm exp}}$ after the release from the trap in the case of weak interactions (quasicondensate limit). The units on both axes are dimensionless. In the trap ($t_{\text{\rm exp}} = 0$) $g_{2}(x) \approx 1$, the asymptotic limit for $t_{\text{\rm exp}}  \rightarrow \infty$ is $g_{2}(x) \approx 1 +
\exp[-2 \left| x \right|/(\lambda_T )]$. }
\label{fig:figure2}
\end{center}
\end{figure}

The theoretical model by Imambekov et al.~\cite{Ima_09} describes the axial shape of the density correlation function $g_{2}(x,t_{\rm exp})$ of a uniform 1D Bose gas as a function of expansion time, 1D density and temperature (compare fig.~\ref{fig:figure2}). To allow for comparison with our experimental finite-size quasicondensate, we determine an averaged two-point correlation function $\tilde{g}_{2}$ defined as follows : We first compute the autocorrelation function for a single integrated density profile $\int n_{\rm 1D}(u) n_{\rm 1D}(u+x) du$ and obtain $G_{2}(x)=\left\langle\int n_{\rm 1D}(u) n_{\rm 1D}(u+x) du\right\rangle$ by averaging over many (several hundred) repetitions. The function $\tilde{g}_{2}$ is then obtained by normalizing with the autocorrelation function of the mean density profile $\int \left\langle n_{\rm 1D}(u) \right\rangle \left\langle n_{\rm 1D}(u+x) \right\rangle du$.

\begin{figure}[tb]
\begin{center}
\includegraphics{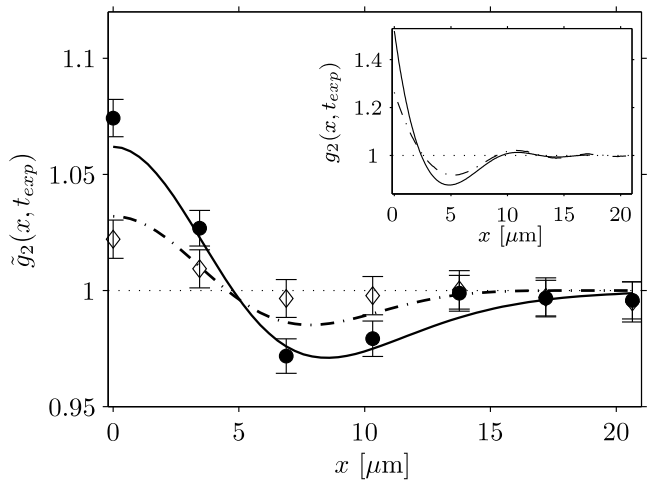}
\caption{Black dots (open diamonds): Measured correlation function $\tilde{g}_{2}(x)$ for
150\,nK (60\,nK) [ i.e., $\hbar t_{\text{\rm exp}}/m \lambda^2_T=0.18 \,(0.03)$]. The solid line (dashed line) shows the according theoretical prediction convolved with the imaging PSF. (Inset) Solid line (dashed line): corresponding theory curves not considering the imaging system.}
\label{fig:figure3}
\end{center}
\end{figure}

Figure~\ref{fig:figure3} shows experimentally obtained two-point density correlation functions for two different temperatures after an expansion time $t_{\rm exp}=10\,$ms.
We observe correlation well in excess of 1 (bunching) for small relative distances, as expected for an only partially coherent source such as a phase fluctuating (and hence multimode) quasicondensate. Furthermore, we find minimum correlations below 1 at finite distance, which is a signature of the transition from the trapped gas $g_{2}(x,t_{\rm exp}=0)\approx 1$ to the far-field limit $g_{2}(x,t_{\rm exp}\rightarrow\infty)\approx 1+\exp(-2 \left| x \right|/\lambda_T)$ throughout the free expansion. 

This behavior can be understood in terms of probability flow in the relative distance space towards $x=0$ resulting in the deficit of probability to find a second atom within a certain distance range.
As $t_{\rm exp}$ increases, this range spreads toward larger $\left| x \right|$. The depth of the minimum decreases as $t_{\rm exp}\rightarrow \infty$, and finally the minimum vanishes for all finite $\left| x \right|$. Note that the surface below the $g_{2}(x,t_{\rm exp})$ function is maintained for all expansion times. Alternatively, this behavior may be described as near-field diffraction in terms of a matter wave analogous to the Talbot effect~\cite{Ima_09}. The two-particle problem is reduced to a single particle one by elimination of the center-of-mass motion. The random phase of the matter wave in the relative coordinate is even due to Bose-Einstein statistics, which leads to a $g_{2}(x)$ peak formation always at $x=0$.
As illustrated in Fig.~\ref{fig:figure2}, the oscillatory behavior will only vanish for expansion times on the order of seconds.

To compare theory and data, the finite resolution of the imaging system has to be taken into account. Typical structure sizes in the density fluctuations are on the order of the system's point spread function (PSF), which hinders the observation of smaller features. The PSF can be determined, starting from a description in Fourier space~\cite{Bue_09}, which includes effects of diffraction, pixel sampling and finite extent of the cloud along the optical axis. It is then approximated by a Gaussian curve with a half width at half maximum (HWHM) of $\sigma_1 = 3.9\,\mu$m. For two-point correlation data, this value becomes $\sigma _2 =\sqrt{2} \sigma _1$. The PSF is finally convolved with the theoretical prediction for the correlation function $g_{2}(x)$. Figure~\ref{fig:figure3} compares a measured $\tilde{g}_{2}(x)$ function to a theoretical calculation based on the experimental parameters and the imaging resolution. We find excellent agreement without any adjustable parameters. 

\begin{figure}[t]
\begin{center}
\includegraphics{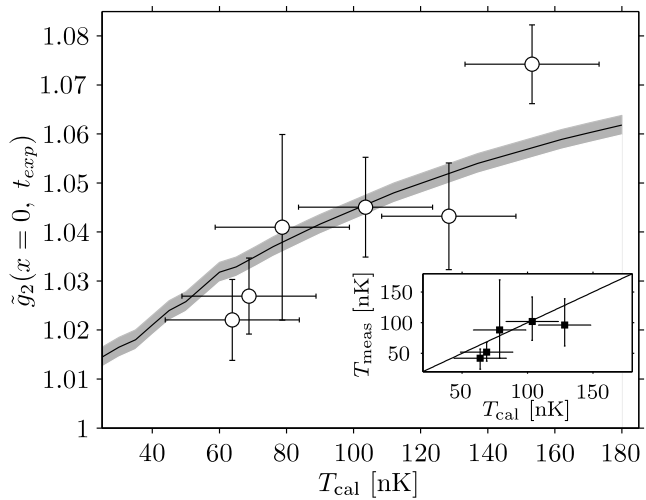}
\caption{Noise-corrected peak height of the correlation function $\tilde{g}_{2}(x=0,t_{exp})$ plotted against the expected temperatures corresponding to the calibration $T_{\rm cal}$. The vertical error bars are given by the statistical fluctuations, the horizontal ones indicate the accuracy of the calibration. The theoretical prediction (grey) 
is plotted for comparison, its width is given by the uncertainty in optical resolution. The inset shows a comparison
of calibrated temperature $T_{cal}$ and measured temperature $T_{\rm meas}$ (black squares) to a line through origin with unity slope (solid).} 
\label{fig:figure4}
\end{center}
\end{figure}

As the expansion time and the atomic 1D density are known to high precision, the density correlation function can be used to determine the temperature of the system in cases where this information cannot be retrieved otherwise. In the interesting temperature range from 20\,nK to the critical temperature $T_c\approx 350\,$nK, the density correlation function at zero distance $g_{2}(0,t_{\rm exp})$ is a monotone function of $T$ that can be calculated in a straightforward fashion from~\cite{Ima_09} and can hence be used for thermometry (compare Fig.~\ref{fig:figure4}).

Before comparing data to theory, again technical aspects of the imaging system have to be considered.
The noise in absorption images is typically composed of atomic and photonic shot noise, and technical detector noise. We study the noise characteristic of our system by evaluating images integrated along the quasicondensate axis with and without atomic signal. It shows that the dominating contribution to noise is technical. Atomic shot noise can be neglected since
the main contribution to the variance on a single pixel arises from the density fluctuations themselves. The technical noise contribution is uncorrelated and therefore contributes to $\tilde{g}_{2}(0,t_{\rm exp})$ only. To account for this noise in the analysis, we use an empty region of each absorption picture and integrate it in the same direction as the region of interest containing the atomic signal. Then we calculate the variance of the empty region and subtract this value from the autocorrelation of the corresponding integrated region of interest before averaging and normalization. 

To evaluate whether measurements of density correlation functions can be used for quantitative comparison, we need an alternative temperature calibration for our samples. For the quasi-1D regime that is currently accessible in our setup, the condensate fraction is between $70\%$ and $90\%$. For higher temperatures ($T>T_c/2$) a more pronounced thermal fraction is present and the temperature of the sample can be determined by fitting a Bose function to the tails of the density profile. For low temperatures, we perform a similar measurement using a special fluorescence imaging scheme~\cite{Bue_09}, which allows us to detect small thermal fractions down to 10$\%$ in a single measurement \footnote{The ability to measure temperatures significantly below $T_c$ is a unique feature of our setup; it does not invalidate a broad interest in alternative thermometry methods based on density correlations.}. To gauge the temperatures of our samples, we relate the temperature of a single measurement to the final value of the RF frequency used for evaporative cooling. We find a linear dependence over a range $4 \ T_c$ to $0.2 \ T_c$ .

In fig.~\ref{fig:figure4} we compare experimental values of $g_{2}(x=0,t_{\rm exp})$ obtained for different temperatures to the theoretical prediction based on~\cite{Ima_09} and find agreement within the error bars. Note that we expect the theory to deviate from the data for temperatures above 110\,nK as the system becomes increasingly three-dimensional. Temperatures obtained from $g_{2}(x=0,t_{\rm exp})$ compared to temperatures derived from the calibration described previously are in agreement with a straight line through origin of slope one. We observe also a deviation from the theoretical prediction for lower temperatures. This is expected since the coherence length approaches the sample length for low temperatures, whereas  theoretical calculations assume an infinite system. 

For our experiments, the stability of the effective trap depth is crucial, since it defines the reference point of the temperature calibration. The precision of $\tilde{g}_{2}(x=0,t_{\rm exp})$ is then given by the statistical fluctuation of the measurement and the noise characteristics of the imaging system. For 300 measurements we achieve an accuracy of $\approx\pm$ 30 nK, where the main limitation is given by the statistical fluctuations.

To conclude, we have studied the two-point density correlation function of quasicondensates in free, collisionless expansion. We find an excess of correlations (bunching) at short inter-particle distance and an oscillatory behavior at finite distances with correlations below unity, in good agreement with the theoretical models of~\cite{Ima_09}. We show that a quantitative comparison can be used for thermometry in regimes, where conventional methods based on absorption imaging fail. We are convinced that two-point density correlations provide a powerful probe for low-dimensional systems, nonequilibrium dynamics~\cite{Hof_07}, and integrability~\cite{Maz_08}.

S.M. and R.B. acknowledge support from the FWF doctoral program CoQuS (W1210-N16). Ch.K. acknowledges funding from the FunMat research alliance. A.P. acknowledges support from the EU (FP7 GA $\rm n^o$~236702). The experimental work was supported by the EU Integrated Project FET/QUIP "SCALA", the FWF project P21080, and the City of Vienna. We thank Dmitry Petrov and Alex Gottlieb for helpfull discussions.

\bibliographystyle{apsrev}

\end{document}